%% file: blood_flow_gps.tex
\documentclass[twocolumn,final]{svjour3}          % twocolumn
\smartqed  % flush right qed marks, e.g. at end of proof
\journalname{Biomechanics and Modeling in Mechanobiology}
\usepackage{amsmath}
\usepackage{amssymb}
\usepackage{graphicx}
\usepackage{color} 
\usepackage{bm}
\usepackage{subfigure}
\usepackage{mathrsfs}
\usepackage{dsfont}
\usepackage{natbib}
\usepackage[misc,geometry]{ifsym}

\begin{document}
\bibliographystyle{spbasic}     
\bibpunct{(}{)}{;}{a}{}{,}

%Title
\title{Generalized Plasma Skimming Model for Cells and Drug Carriers in the Microvasculature}

%Author
\author{Tae-Rin Lee \and
Sung Sic Yoo \and Jiho Yang
}

%Institute
\institute{ Tae-Rin Lee (\Letter) \and Sung Sic Yoo \and Jiho Yang
            \at Advanced Institutes of Convergence Technology, Seoul National University, Suwon 443-270, Republic of Korea\\
            e-mail: taerinlee@snu.ac.kr
            \and
            Jiho Yang
            \at Department of Computer Science, Technische Universit{\"a}t M{\"u}nchen, Boltzmannstra{\ss}e 3, Garching, Germany
}

\date{Received: date / Accepted: date}

\maketitle

%Abstract
\input{abstract}

%Keyword
\keywords{Plasma skimming \and Microvascular transport \and Drug carrier \and Nanomedicine}

%Sections
\input{introduction}

\input{materials_and_methods}

\input{results}

\input{discussion}

%Acknowledgement
\section*{Acknowledgment}
This research was supported by Basic Science Research Program through the National Research Foundation of Korea (KRF) funded by the Ministry of Education (NRF-2015R1D1A1A01060992), and by the Bio \& Medical Technology Development Program of the National Research Foundation (NRF) funded by the Ministry of Science, ICT \& Future Planning (2016M3A9B4919711).

\section*{Conflict of Interest}
The authors declare that they have no conflict of interest.

%References
\bibliography{blood_flow}  

\end{document}

%% file: abstract.tex
\begin{abstract}

In microvascular transport, where both blood and drug carriers are involved,
plasma skimming has a key role on changing hematocrit level and drug carrier
concentration in capillary beds after continuous vessel bifurcation in the
microvasculature. While there have been numerous studies on modeling the plasma
skimming of blood, previous works lacked in consideration of its interaction
with drug carriers. In this paper, a generalized plasma skimming model is
suggested to predict the redistributions of both the cells and drug carriers at
each bifurcation. In order to examine its applicability, this new model was
applied on a single bifurcation system to predict the redistribution of red
blood cells and drug carriers. Furthermore, this model was tested at
microvascular network level under different plasma skimming conditions for
predicting the concentration of drug carriers. Based on these results, the
applicability of this generalized plasma skimming model is fully discussed and
future works along with the model's limitations are summarized.

\end{abstract}

%% file: introduction.tex
\section{Introduction}

Drug carriers with chemotherapeutic agents are administrated to
macrocirculatory systems for efficient diagnosis and treatment of various
diseases~\citep{Jain-Nanomedicine,Langer-Nanocarrier}. In order to deliver drug
carriers in diseased areas, the geometric difference between normal and
diseased areas of microvasculatures is mainly
considered~\citep{Jain-Normalization}. For this reason, good understanding of
the physics of drug carriers in the microvasculature holds significant
importance. Related to this topic, several experiments were performed to
quantify the drug carrier distributions  \textit{in vitro} \citep{Exp-NP-Disp,
tan2016characterization} and \textit{in vivo}
\citep{Jain-Normalization,Munn-Simultaneous,Van-Rapid}. Even though these
experimental works suggested good methods at single vessel and single
bifurcation level, quantitative measurement of the drug carrier concentration
in the entire microvasculature still remains as a challenge. 

An alternative way of predicting the drug carrier distribution is to develop
a mathematical model that can estimate the targeting
efficiency of drug carriers in the microvasculature. Several computational
methods were suggested and conducted to simulate various interactions between cells and drug
carriers \citep{TR-NP-dispersion,TR-NP-UQ, Fedosov-SciRep,
tan2016characterization}. Although the blood flow simulations with drug
carriers in a microvessel are possible, Computational Fluid Dynamics (CFD) simulation of cells and
drug carriers \textit{in the entire microvasculature} has not been reported yet
due to its tremendous efforts for a test-run. Meanwhile, mathematical models
have been utilized to simplify the blood flow in microvessels by applying
Poiseuille's law with the \textit{in vivo} viscosity law \citep{Pries4} to
calculate the blood flow in microvessels, along with conservation of mass for
flow division at each bifurcation. By using these mathematical models, drug
carrier distributions in the tumor microvasculature were computed at
microvascular-network level \citep{frieboes2013computational,
curtis2015computational}. 

\begin{figure}[b]
\centering
\includegraphics[scale=0.9]{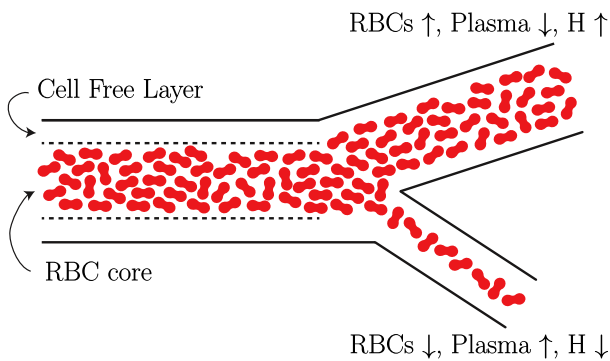} 
\caption{Plasma skimming of RBCs at a bifurcation.}
\label{fig1}
\end{figure}

Previous mathematical models were focused on the plasma
skimming \citep{krogh1922studies, Pries4} of blood flow at each bifurcation. In
the microvascular transport, red blood cells (RBCs) are concentrated on the
vessel core with cell-free layer (CFL) near the vessel wall.  Due to these two
areas, hematocrit level is changed in following daughter vessels. The
subsequent bifurcating processes change the total distribution of RBCs and
blood flow characteristics in the entire microvasculature, and this is called
the plasma skimming effect. However, the drug carrier distribution in the
microvasculature is highly dependent on the plasma skimming of blood flow.
Hence, it is important to develop a mathematical model that can predict
plasma skimming effects of both blood and drug carriers considering their
interactions. 

Previous studies showed that their interactions depend on numerous factors
including drug carriers' size, shape, and even vessel geometry
\citep{tan2013influence}. Both numerical simulations and experiments indicated
clear interactions between drug carriers and RBCs \citep{Fedosov-SciRep,
d2016microfluidic}. Lee et al. studied that the cross-sectional distribution of
drug carriers depend on their size, and that larger particles tend to marginate
towards vessel walls \citep{TR-NP-dispersion}. On the other hand, Muro et al.
and also Namdee et al. observed that larger particles can result in lower
targeting efficiency \citep{muro2008control, namdee2014vivo}. These findings
highlight that whilst drug carriers and RBCs clearly interact with great
significance, there is no absolute rule to describe how they interact and how
drug carriers distribute \citep{champion2008role, sobczynski2015drug}.
Therefore, it is necessary to study the drug carrier distribution and its
delivery by coupling it with plasma skimming and the resulting interactions.
Moreover, this must be conducted in a generalized fashion and also in simplest
form for computationally efficient prediction. In this paper, a generalized
plasma skimming model for cells and drug carriers in the microvasculature is
suggested. To achieve this goal, the recently developed plasma skimming model
for blood in microcirculatory networks \citep{GOULD} is applied, and extended to
the scope of both cells and drug carriers. Two cases with different drug carrier
distribution condition in a single bifurcation model are considered to validate
the model.  Furthermore, the generalized plasma skimming model is tested at
microvascular network level for estimating distributions of cells and drug
carriers. In addition, the limitation of current mathematical model and future
perspectives are briefly discussed. 

\begin{figure*}
\centering
\includegraphics[scale=0.45]{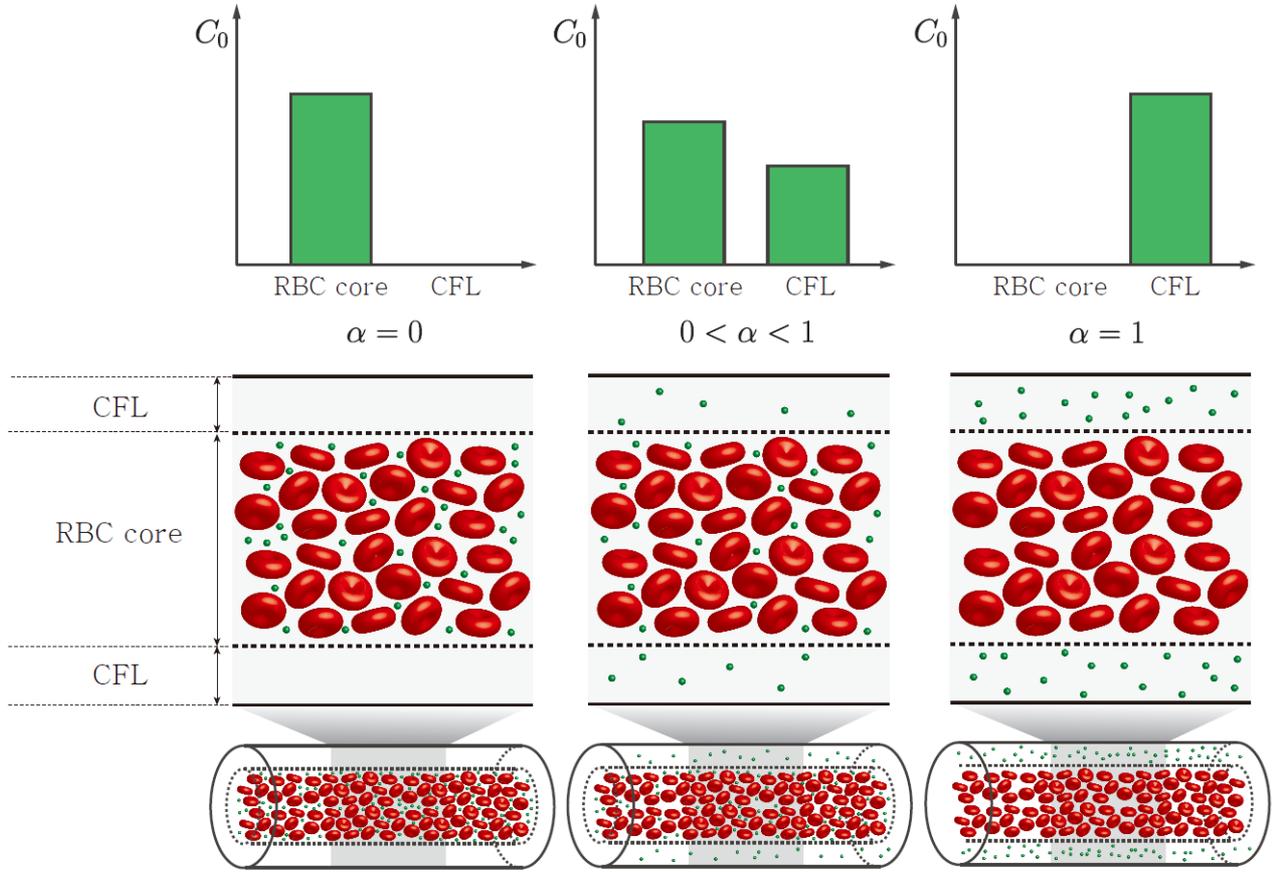}
\caption{Cross-sectional distributions of RBCs and drug carriers for generalized plasma skimming model. $\alpha$ is the relative distribution between RBCs and drug carriers. At $\alpha$ = 0, all drug carriers remain in the RBC core. When $\alpha$ is increased, drug carrier distribution shifts toward CFL.}
\label{fig2}
\end{figure*}

%% file: materials_and_methods.tex
\section{Materials and Methods}

%------------------------------------------------------------------------------------Plasma skimming model for blood
\subsection{Plasma skimming model for blood}
In a parent vessel, blood flow is determined by blood viscosity, hematocrit,
vessel geometry, and pressure drop. After bifurcation, the blood flow
characteristics in the following daughter vessels are changed due to their
diameter reduction. The flow rate in the daughter vessel can be easily
calculated by using conservation of mass with the diameters of daughter vessels. In
the parent vessel, RBCs are concentrated on the core of microvessel with a
certain thickness of CFL. Then, the thinner daughter vessel, as a side branch,
takes less RBCs and more plasma. On the other hand, the thicker daughter vessel
takes more RBCs and less plasma. As shown in Fig.~\ref{fig1}, the thicker
daughter vessel has higher hematocrit (hemoconcentration) than the parent
vessel and the thinner vessel has lower hematocrit (hemodilution). This
phenomenon is called the ``plasma skimming". To  model the plasma skimming
parametrically, \citet{GOULD} suggested a drift parameter, $M$. The general
steps to determine hematocrit changes in a bifurcation are as follows: 

\begin{equation}
H_1 = H_0 - \Delta H = \zeta^{b}_1 H^* 
\label{eq:bPS1}
\end{equation}
\begin{equation}
H_2 = \zeta^{b}_2 H^*
\label{eq:bPS2}
\end{equation}
\begin{equation}
Q_0 H_0 = Q_1 H_1 + Q_2 H_2 = Q_1 \zeta^{b}_1 H^* + Q_2 \zeta^{b}_2 H^*
\label{eq:bPS3}
\end{equation}
\begin{equation}
\zeta^{b}_i = \left( \frac{A_i}{A_0} \right) ^{\frac{1}{M}} \\\ \textrm{where}\ i = 1, 2
\label{eq:bPS4}
\end{equation}

\noindent where $H$ is the hematocrit, $\zeta^b$ is the hematocrit change
coefficient due to plasma skimming, $Q$ is the flow rate, $A$ is the
cross-sectional area of each vessel, and subscription 0, 1 and 2 indicate the
parent, and two daughter vessels, respectively. If two daughter vessels have
the same vessel diameter, $\zeta^b$ values for determining hematocrit of daughter
vessels are also identical such that the hematocrit values of daughter vessels
are the same with that of parent vessel. In most cases, however, two daughter
vessels have different diameters. If the diameters of two daughter vessels
greatly differ, the thicker daughter vessel has higher $\zeta^b$ and the
thinner daughter vessel has lower $\zeta^b$. Note that while $\zeta^b$ is a
dimensionless kinematic quantity relating the relative phase velocities in
terms of the ratio of parent to daughter vessel cross-sectional areas, the
absolute number of $\zeta^b$ does not describe plasma skimming. Only the
relative difference between $\zeta^b$ of the daughter vessels affects the
hematocrit of daughter vessels. Whilst there are other mathematically simple
and validated plasma skimming models \citep{pries2005microvascular,
guibert2010new}, the model developed by Gould et al. \citep{GOULD} was considered due to its
easy extensibility. 

\begin{figure*}
\centering
\includegraphics[scale=1.3]{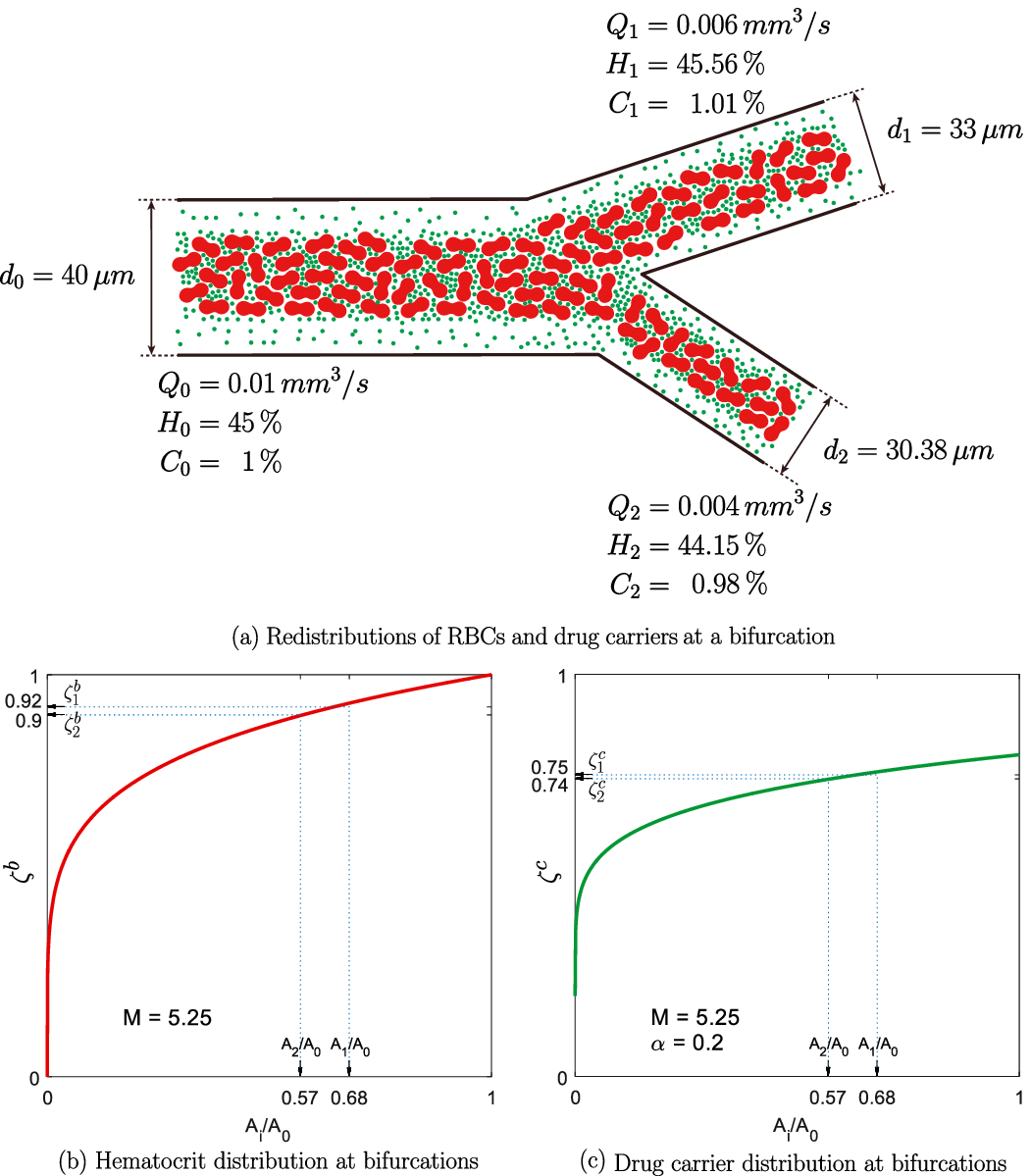}
\caption{Redistribution of RBCs and drug carriers with $\alpha = 0.2$ at a bifurcation. Low $\alpha$ indicates drug carriers stay in the RBC core of parent microvessel. After the bifurcating process, RBCs and drug carriers are redistributed in the daughter vessels. (a) Schematic diagram of single bifurcation with RBCs and drug carriers. (b) $\zeta^b$ curve for calculating hematocrit in daughter microvessels. (c) $\zeta^c$ curve for calculating concentration of drug carriers in daughter microvessels.}
\label{fig3}
\end{figure*}

\begin{figure*}
\centering
\includegraphics[scale=1.3]{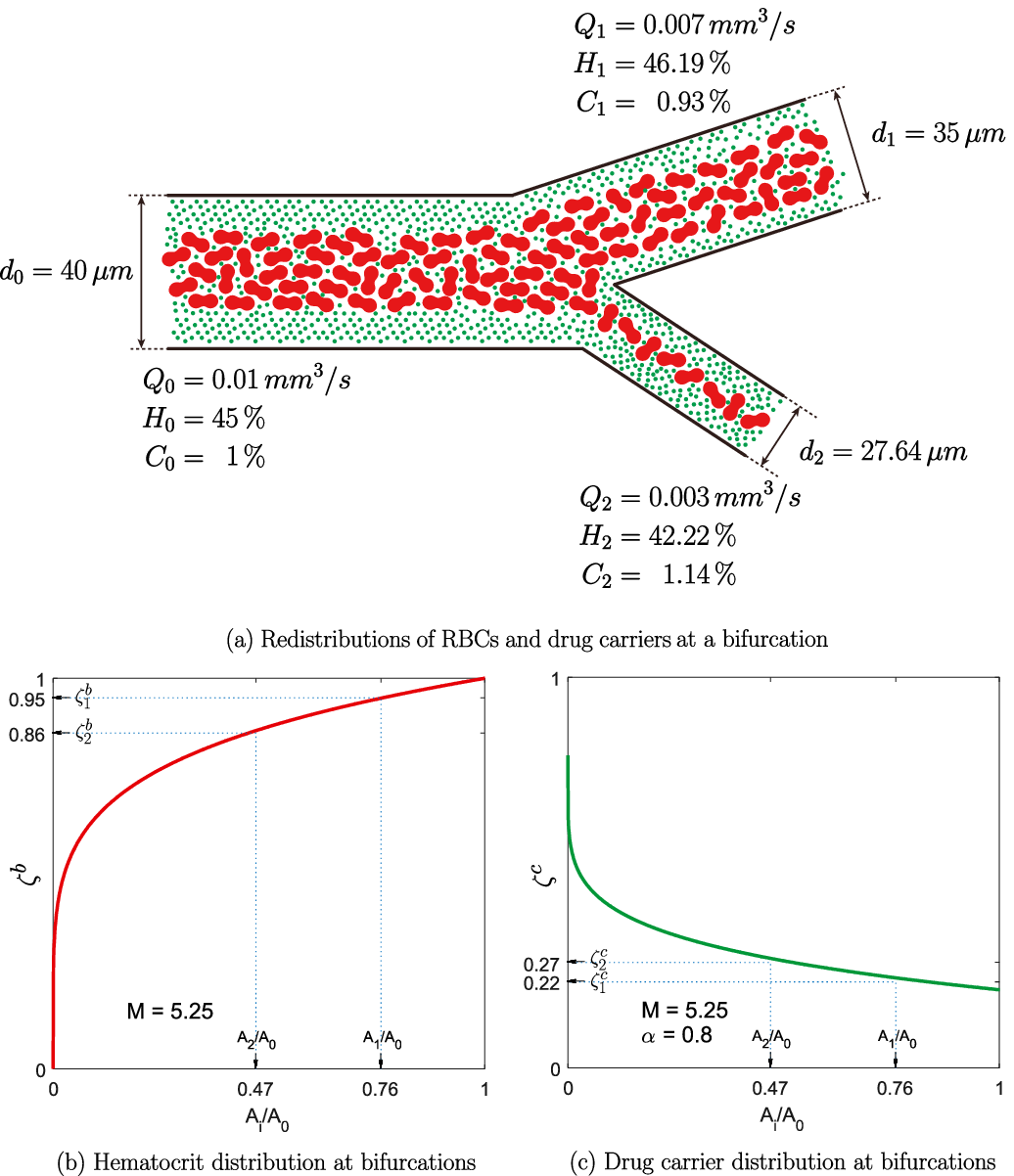}
\caption{Redistribution of RBCs and drug carriers with $\alpha = 0.8$ at a bifurcation. High $\alpha$ indicates drug carriers stay in the CFL area of parent microvessel. After the bifurcating process, RBCs and drug carriers are redistributed in the daughter vessels. (a) Schematic diagram of single bifurcation with RBCs and drug carriers. (b) $\zeta^b$ curve for calculating hematocrit in daughter microvessels. (c) $\zeta^c$ curve for calculating concentration of drug carriers in daughter microvessels.}
\label{fig4}
\end{figure*}

\begin{figure*}
\centering
\includegraphics[scale=0.7]{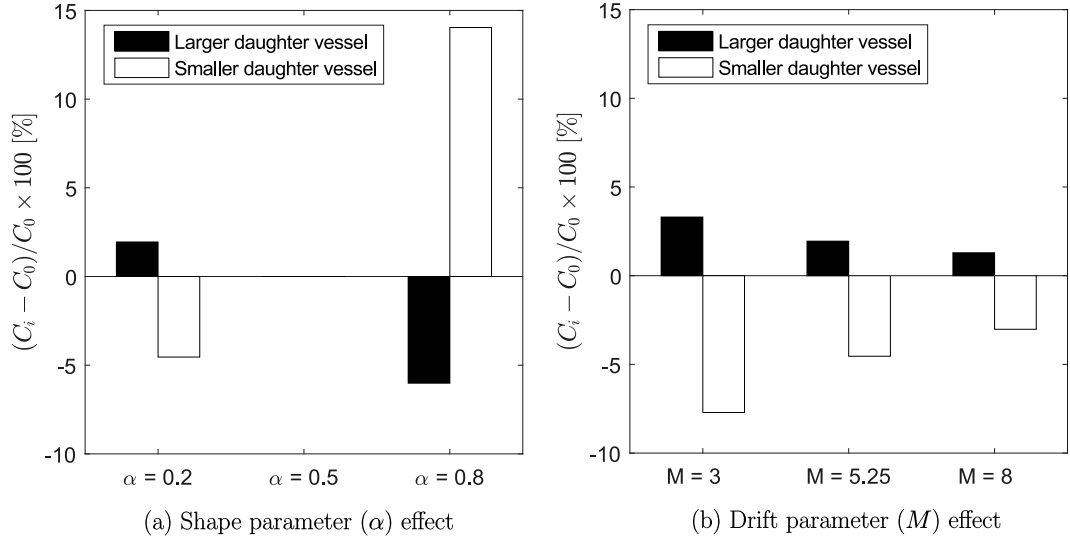}
\caption{Effects of $\alpha$ and $M$ in a single bifurcation model. The average changes in drug carrier distribution were calculated for a single bifurcation model with different $\alpha$ and $M$. (a) Three $\alpha$ values were tested. When $\alpha$ = 0.5, drug carriers in the parent vessel were uniformly distributed throughout the vessel's cross-section, hence not producing any change in drug carrier redistribution. With $\alpha < 0.5$, since drug carriers are highly accumulated in the RBC core, the concentration of drug carriers in the larger daughter vessel was increased. On the other hand, for $\alpha > 0.5$, the concentration of drug carriers in the smaller daughter vessel was increased up to 14 \%. (b) When $\alpha$ = 0.2, the drift parameter, $M$, effect was investigated by applying three $M$ values. When $M$ is low, the concentration of drug carriers in the daughter vessels was significantly changed.}
\label{fig5}
\end{figure*}

\begin{figure*}
\centering
\includegraphics[scale=0.9]{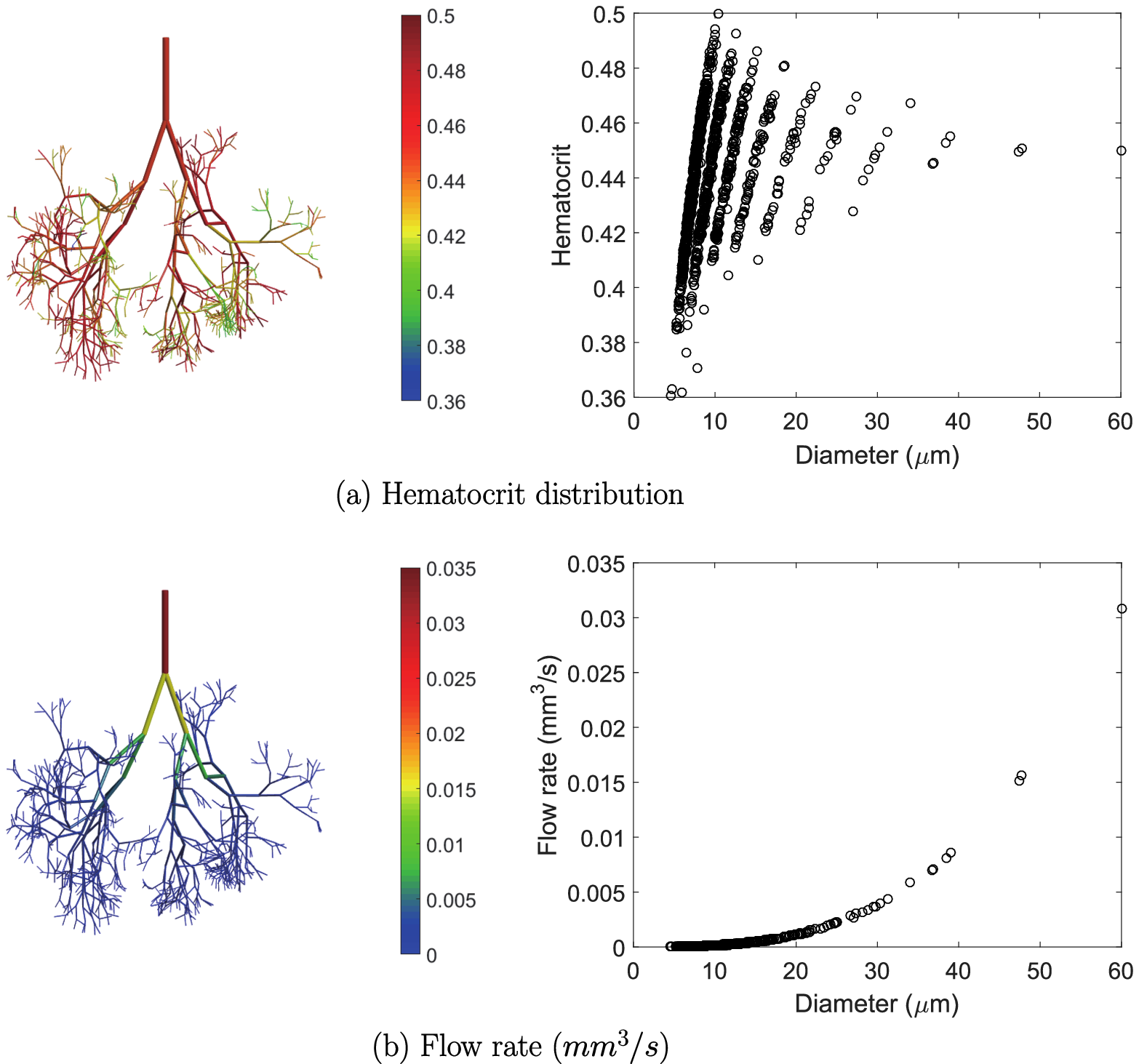}
\caption{Representative model of microvascular transport in a microvascular network. The microvasculature geometry was constructed by a computational algorithm with model parameters, $\gamma$ = 3 and $\beta$ =  25 with root vessel diameter = 60 $\mu m$. The pressure drops between the root vessel and the capillary ends were 60 mmHg. (a) Hematocrit distribution along diameter. The hematocrit distribution was calculated by \citet{GOULD}'s method. (b) Flow rates along diameter. The blood flow in the microvessels was calculated by Poiseuille's law with the \textit{in vivo} viscosity law \citep{Pries4}.}
\label{fig6}
\end{figure*}

%------------------------------------------------------------------------------------Generalized plasma skimming model for cells and drug carriers
\subsection{Generalized plasma skimming model for cells and drug carriers}
The changes in hematocrit due to the plasma skimming are a special case of mass
transport at a bifurcation. In order to mathematically model the distribution
of drug carriers or any cells, which may not necessarily be the same as the one
of RBCs, a generalized plasma skimming model was constructed by adding a
distribution-shape parameter $\alpha$ to the plasma skimming model of blood
\citep{GOULD} as follows:

\begin{equation}
C_1 = C_0 - \Delta C = \zeta^c_1 C^* 
\label{eq:gPS1}
\end{equation}
\begin{equation}
C_2 = \zeta^c_2 C^*
\label{eq:gPS2}
\end{equation}
\begin{equation}
Q_0 C_0 = Q_1 C_1 + Q_2 C_2 = Q_1 \zeta^c_1 C^* + Q_2 \zeta^c_2 C^*
\label{eq:gPS3}
\end{equation}
\begin{equation}
\zeta^{c}_i = (1-\alpha) \left( \frac{A_i}{A_0} \right) ^{\frac{1}{M}}  + \alpha \left\{1- \left(\frac{A_i}{A_0}\right)^{\frac{1}{M}}\right\}  \\\ ~~\textrm{where}\ i = 1, 2
\label{eq:gPS4}
\end{equation}

\noindent where $C$ is the concentration of cells or drug carriers, $\zeta^c$
is the concentration change coefficient and $\alpha$ is the shape parameter for
cross-sectional distribution of $C$ in a microvessel, as described in
Fig.~\ref{fig2}. In the case of $\alpha = 0$, all cells or drug carriers are
within RBC region, equivalent to the original plasma skimming model for RBCs.
On the other hand, when $\alpha = 1$, all cells or drug carriers are within CFL
region. They are redistributed by the opposite tendency of plasma skimming. In
most cases $\alpha$ is likely to be somewhere between 0 and 1. For example, in
a previous work \citep{TR-NP-dispersion}, it was reported that larger particles
stay in CFL region and smaller particles stay in RBC core. In this case,
particle size is proportional to $\alpha$. Therefore, in such case, $\alpha$
effectively becomes an indicator to quantify the transport efficiency in
capillary beds. In
order to simplify the model, cells or drug carriers were assumed to be
homogeneous in both RBC and CFL regions with $\alpha$ representing the relative
level of cells or drug carriers in the two regions. The shape parameter
$\alpha$ enables the modeling of drug carrier redistribution after bifurcation
tailored to the characteristics of specific drug carriers, and hence making the
model to be very generalized. Therefore, if $\alpha$ of a specific drug carrier
is studied, this mathematical model can be adjusted on it.

\begin{figure*}
\centering
\includegraphics[scale=1.0]{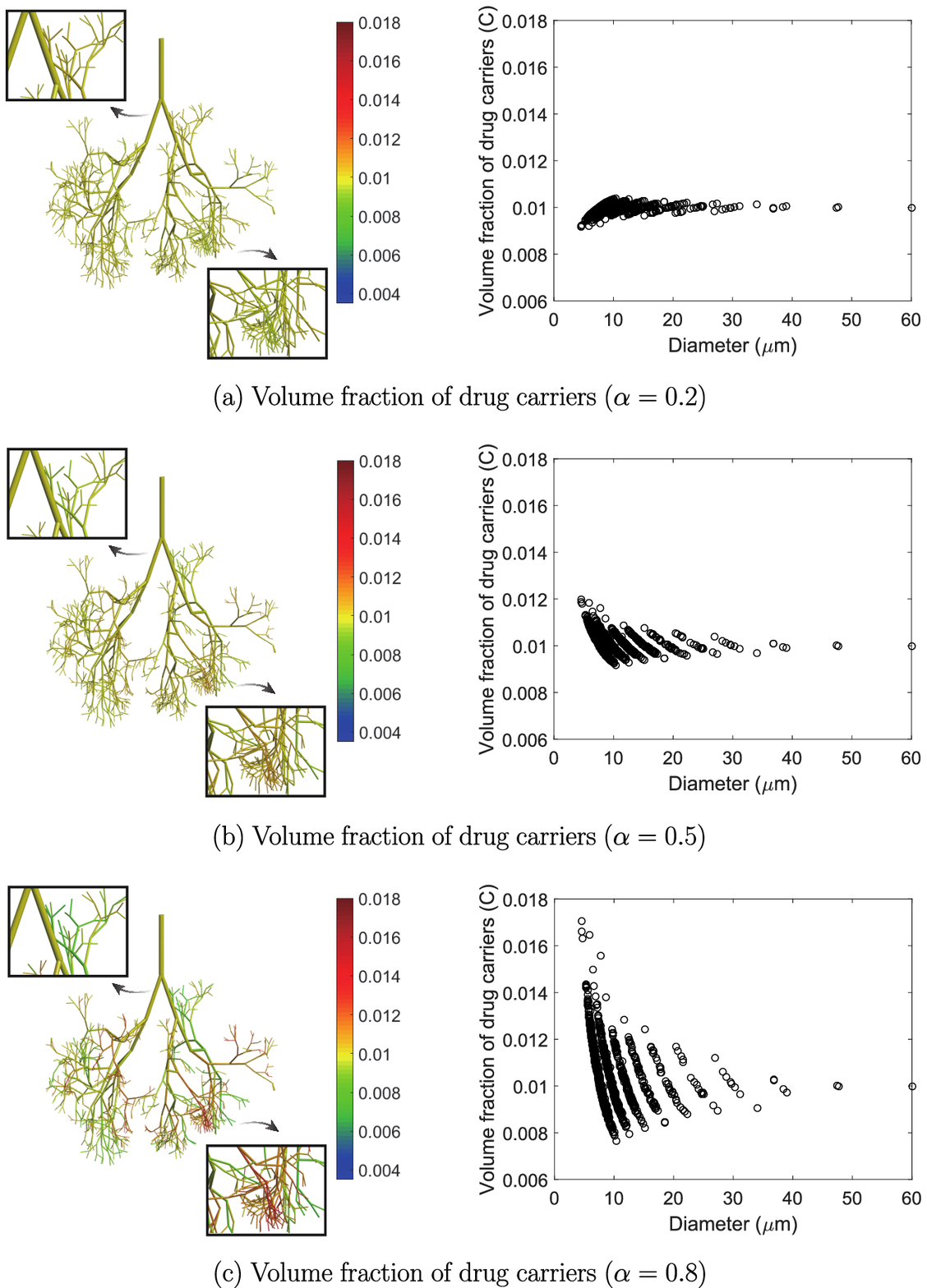}
\caption{Redistribution of drug carriers in the microvasculature. The drug carrier distribution was calculated by Eqs.~\eqref{eq:gPS1}-\eqref{eq:gPS4}. Three $\alpha$ values were used to predict the redistribution tendency of drug carriers in the computational microvasculature model. The initial concentration of drug carriers in the root vessel was 0.01. By increasing $\alpha$, the concentration of drug carriers in the capillary beds, below 10 $\mu$m diameters, was in the range from 0.008 to 0.017, and the average concentration in the capillary beds was increased.}
\label{fig7}
\end{figure*}

%------------------------------------------------------------------------------------Coupling the generalized plasma skimming model with blood flow in the microvasculature
\subsection{Coupling the generalized plasma skimming model with blood flow in
the microvasculature}\label{method_microvasculature} Computing drug carrier
distribution by using the generalized plasma skimming model is simple in a
single bifurcation case. However, it is challenging to predict distributions of
cells and drug carriers in the entire microvascular network. If one is to
consider individual RBCs and drug carriers in a microvascular network with
thousands of vessel segments, it requires tremendous amount of computation time
with heavy parallel processing. Furthermore, \textit{in vivo} experiments are
still challenging to accurately measure the distributions. In order to greatly
reduce the computation time while obtaining feasible solutions, the generalized
plasma skimming model was coupled with the mathematical model of blood flow to
predict the distributions. The microvascular network geometry was
computationally generated based on mathematical algorithms, not from scanned
images, by choosing vessel diameter ($d_i$), vessel length ($l_i$) and
bifurcation angles ($\theta_i$ and $\phi_i$) \citep{yang2016predicting}. Three
diameters for each bifurcation (one parent vessel and two daughter vessels)
were governed by $d^{\gamma}_0=d^{\gamma}_1 + d^{\gamma}_2$  where $\gamma$ was
fixed at 3 \citep{sherman, Murray}. For defining the vessels' lengths from
their diameters, the length-to-diameter ratio ($\beta = l / d$) was fixed at 25
\citep{yang2016predicting}. Flow rates of blood ($Q_i$) were calculated by
Poiseuille's law and the \textit{in vivo} viscosity \citep{Pries4}:

\begin{equation}
Q_i = \frac{\pi d_i^4}{128\mu_i l_i} \Delta P_i\label{eq:flow_rate1}
\end{equation}
\begin{equation}
\begin{split}
\mu_{i} = \mu_p\biggl[1+(\mu_{0.45}^*-1)\frac{(1-H_D)^C-1}{(1-0.45)^C-1} \\
\times \biggl(\frac{d}{d-1.1}\biggl)^2\biggl]\biggl(\frac{d}{d-1.1}\biggl)^2 \label{eq:in_vivo_viscosity}
\end{split}
\end{equation}
\begin{equation}
\mu^*_{0.45} = 6e^{-0.085d}+3.2-2.44e^{-0.06d^{0.645}}
\label{eq:in_vivo_viscosity2}
\end{equation}
\begin{equation}
\begin{split}
C = (0.8+e^{-0.075d})\biggl(-1+\frac{1}{1+10^{-11}d^{12}}\biggl) \\
+ \frac{1}{1+10^{-11}d^{12}}.
\label{eq:in_vivo_viscosity3}
\end{split}
\end{equation}

\noindent Here, $\mu_i$ is the apparent viscosity of blood and $\mu_p$ is the
reference viscosity of plasma, fixed at $9\cdot 10^{-6}$ mmHg$\cdot$s
\citep{YANG}. $H_D$ is the discharge hematocrit, which was calculated by Eqs.
\eqref{eq:bPS1}, \eqref{eq:bPS2}, \eqref{eq:bPS3} and \eqref{eq:bPS4}. Also,
three flow rates at each bifurcation were calculated by conservation of mass,
$\sum Q_i = 0$. By assuming that the microvascular transport of drug carriers
is independent of the blood flow, the drug carrier distribution can be
calculated by explicitly coupling blood flow rates in microvessels with the
generalized plasma skimming model. Note that the drift parameter, $M$, is
identical with that for hematocrit calculation, and the proposed model is fully
coupled with hematocrit and plasma skimming of RBCs.

%% file: results.tex
\section{Results}

%---------------------------------------------------------------------------------Plasma skimming of drug carriers in a single bifuraction model
\subsection{Plasma skimming of drug carriers in a single bifurcation model} 
The single bifurcation model used for testing the plasma skimming effects is
given in Fig.~\ref{fig3}a. The diameters of a parent and two daughter vessels
were 40, 33 and 30.38 $\mu$m, respectively. The flow rates of three
microvessels were 0.01, 0.006 and 0.004 $mm^3/s$, respectively. In the parent vessel, the
hematocrit ($H_0$) and the concentration of drug carriers ($C_0$) were given as
45 \% and  1 \% of whole blood volume. In microvascular transport, the
concentration of RBCs is increased in the larger daughter vessel and decreased
in the smaller daughter vessel due to the plasma skimming effect. The
hematocrit changes in two daughter vessels were calculated by Eqs.
\eqref{eq:bPS1}-\eqref{eq:bPS4}. The drift parameter, $M$ for \textit{in vivo}
microvascular network was fixed at 5.25 as previously reported \citep{GOULD}.  

From the given conditions, hematocrit changes followed the curve in
Fig.~\ref{fig3}b. The hematocrit in larger daughter vessel was increased to
45.56 \% and the hematocrit in smaller daughter vessel was decreased to 44.15
\%. In a single bifurcation model, the concentration changes in drug carriers
in two daughter vessels are predicted by Eqs. \eqref{eq:gPS1}-\eqref{eq:gPS4}.
As shown in Fig.~\ref{fig2}, drug carriers are concentrated on the RBC core at
$\alpha$ = 0.2. Therefore, the concentration changes in drug carriers should
follow the same trend of blood hematocrit. The concentration in the larger
daughter vessel, $C_1$ was 1.01\% (increased) and the concentration in the
smaller daughter vessel, $C_2$ was 0.98 \% (decreased) (Fig.~\ref{fig3}c),
hence showing the expected tendency. 

Figure~\ref{fig4}a shows the redistribution of RBCs and drug carriers at
$\alpha$ = 0.8. Initial concentration of drug carriers was low in RBC core and
high in CFL, which means drug carriers tend to easily escape from RBC core and
marginate toward CFL. Then, the plasma skimming at the bifurcation caused drug
carriers to accumulate in the smaller daughter vessel, or in better words, in
the vessel where there is more CFL. Consequently, the concentration of drug
carriers was decreased in larger daughter vessel and
increased in smaller daughter vessel. The hematocrit changes, $\zeta_b$, are
proportional to the diameter changes, $A_i/A_0$, as plotted in
Fig.~\ref{fig4}b. On the other hand, the concentration changes in drug
carriers, $\zeta^c$, are inversely proportional to $A_i/A_0$, as described in
Fig.~\ref{fig4}c. The concentration of drug carriers in the larger daughter
vessel, $C_1$, was decreased to 0.93 \% compared to $C_0$ = 1 \%. Moreover, the
concentration of drug carriers in the smaller daughter vessel, $C_2$, was
increased to 1.14 \%. 

The effect of shape parameter on drug carrier redistribution is summarized in
Fig.~\ref{fig5}a. The initial and boundary conditions of single bifurcation
were the same with that in Fig.~\ref{fig4}a. Note that the concentration
changes at $\alpha$ = 0.5 were zeros for both daughter vessels because the drug
carriers were uniformly distributed in the microvessel. At $\alpha$ = 0.2, the
concentration of drug carriers was increased by 2 \% in the larger
daughter vessel and decreased by 5 \% in the smaller daughter vessel. On the
other hand, at $\alpha$ = 0.8, the concentration of drug carriers was decreased
by 7 \% in the larger daughter vessel and increased by 14 \% in the smaller
daughter vessel. In Fig.~\ref{fig5}b, the effect of drift parameter, $M$, is
investigated at $\alpha$ = 0.2. High $M$ implies well-mixed RBCs and plasma.
Low $M$ implies highly separated RBC core and CFL. The model was tested with
three cases at $M$ = 3, 5.25 and 8, and the concentration changes in drug
carriers showed increasing tendency for decreasing $M$. Intuitively, at $M$ = 3
or lower, separated RBC core and CFL regions change the distribution of drug
carriers. At $M$ = 8, the well-mixed blood makes the distribution of drug
carriers to stabilize.

%---------------------------------------------------------------------------------Generalized plasma skimming of drug carriers in the microvasculature
\subsection{Generalized plasma skimming of drug carriers in the microvasculature}\label{gps}
The generalized plasma skimming model for RBCs and drug carriers was applied to
predict their distributions in the microvascular network. The mathematically
generated computational microvasculature model mentioned in Section
\ref{method_microvasculature} was used. Note that this computational
microvasculature model is limited for generating binary tree-like structures.
However, \textit{in vivo} capillary beds form loops, anastomoses, and
multifurcations. Therefore, for a more realistic analysis it is necessary to
consider a true network of microvessels rather than just continuously
bifurcating microvessels. Despite such limitation, this binary tree-like
structure is sufficient for studying drug carrier redistribution in
continuously branching microvessels. For generating a more realistic
microvascular network, one can refer to the computational microvasculature
constructed by \cite{linninger2013cerebral}. From the root vessel with 60
$\mu$m diameter, the vessel segment was bifurcated ten times until the
diameters were decreased to under 10 $\mu$m, with $\beta$ = 25
\citep{yang2016predicting}. The boundary condition was 60 mmHg pressure drops
between inlet root vessel and outlet capillary ends. For calculating hematocrit
distributions, the drift parameter, $M$, was fixed at 5.25 \citep{GOULD}. The
initial hematocrit in the root vessel was 0.45. Figure~\ref{fig6} shows the
hematocrit distributions and flow rates in the given microvascular network.
Unlike the initial hematocrit in the root vessel, the hematocrit distribution
was significantly changed due to plasma skimming. In capillary
ends, blood hematocrit was in the range from 0.35 to 0.5. Based on the
hematocrit and \textit{in vivo} viscosity law, flow rates of blood were
calculated and plotted along with diameter in Fig.~\ref{fig6}b. Furthermore,
the volumetric flow rate was proportional to vessel diameter, as plotted in
Fig.~\ref{fig6}b. 

\begin{figure}
\centering
\includegraphics[scale=0.5]{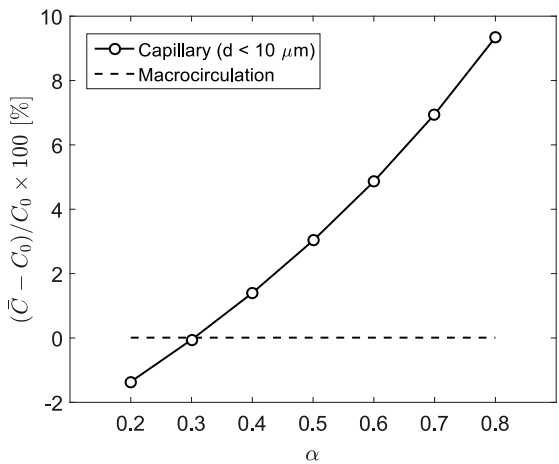}
\caption{Effect of $\alpha$ in a microvascular system. The average concentration of drug carriers with different $\alpha$ values was predicted in the microvascular system. The initial concentration of drug carriers in the root vessel was assumed to be the same as the one in the macrocirculation.}
\label{fig8}
\end{figure}

\begin{figure*}[h]
\centering
\includegraphics[scale=0.83]{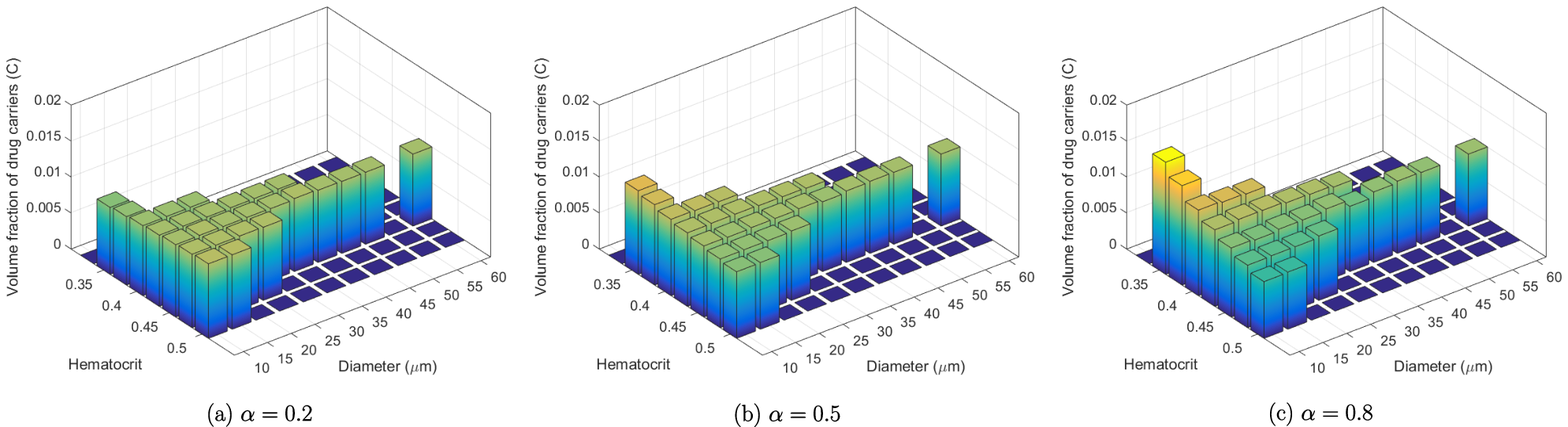}
\caption{Effect of $\alpha$ along with vessel diameter and hematocrit in a microvascular system. The concentration of drug carriers are plotted along with hematocrit and vessel diameter. Three $\alpha$ cases were tested in the computational microvascular network model.}
\label{fig9}
\end{figure*}

In the microvascular transport of blood in the microvasculature, the
distribution of drug carriers was predicted by
Eqs.~\eqref{eq:gPS1}-\eqref{eq:gPS4}. The initial volume fraction of drug
carrier was 0.01. Three cross-sectional distributions of drug carriers,
$\alpha$ = 0.2, 0.5 and 0.8, were applied to the computational microvasculature
model. Figure~\ref{fig7} shows the results at $\alpha$ = 0.2, 0.5
and 0.8. When $\alpha$ was higher, the drug carriers were accumulated in the
capillary beds with high variances, ranging from 0.008 to 0.017. From these
results, it is expected that highly accumulated drug carriers in
CFL of large root vessels can move to capillary ends quickly by following the
plasma skimming of blood. For quantitatively evaluating the average
accumulation of drug carriers in the capillary ends, an average volume fraction
is defined as:

\begin{equation}
\bar C = \frac{1}{n_{cap}}\sum_{i=1}^{n_{cap}}C_i
\end{equation}

\noindent where $\bar C$ is the average volume fraction of drug carriers and
$n_{cap}$ is the total number of microvessels with diameters below 10 $\mu$m.
The relative variance of volume fraction is $(\bar C-C_0)/C_0 \times 100$ \%
where $C_0$ is the initial volume fraction of drug carriers. Note that $C_0$ =
0.01 (or 1\%). Figure~\ref{fig8} shows the changes in average volume fraction
for the $\alpha$ range from 0.2 to 0.8. Interestingly, when $\alpha$ was
increased, the average accumulation of drug carriers in capillary ends was
quasi-linearly increased up to 10\% of initial volume fraction. Figure~\ref{fig9}
shows the redistribution of drug carriers in the microvascular network with
respect to hematocrit and diameter. At $\alpha$ = 0.2, the volume fraction of
drug carriers was slightly changed. However, as plotted in Fig.~\ref{fig9}c, in
the case of low hematocrit and small diameter vessel at $\alpha$ = 0.8, the
volume fraction of drug carriers was greatly increased. 

%% file: discussion.tex
\section{Discussion}

The generalized plasma skimming model was developed for predicting the
distributions of cells and drug carriers in the microvasculature. As expected,
the concentration of drug carriers in the microvasculature was highly changed
by the plasma skimming of blood. Firstly, in a single bifurcation, the
redistributions of RBCs and drug carriers were calculated at two $\alpha$
values, which represent RBC core concentrated drug carriers ($\alpha$ = 0.2)
and CFL concentrated drug carriers ($\alpha$ = 0.8). Secondly, the generalized
plasma skimming model was applied to predict the concentration of drug carriers
in the microvasculature. Furthermore, the redistribution of drug carriers was
correlated with hematocrit and their cross-sectional distribution in the entire
microvasculature. Therefore, if the specific parameters of microvascular
environment are given for targeting diseased areas, the design of drug carriers
can be suggested by this mathematical model, specifically by selecting
cross-sectional distribution of drug carriers, called shape parameter,
$\alpha$. 

Related to this topic, \citet{thomas2014characterization} fully discussed about
the effect of vessel bifurcation in delivering nanoparticles with different
sizes. From the experiments, it was found that the drug delivering efficiency
was highly related to the microvascular environment. However, it is still
challenging to quantitatively measure the concentration of drug carriers
\textit{in vivo} mouse models. Also, due to the limited information from
experiments, it is difficult to decide the design of drug carriers for
targeting specific microvascular environments. Hence, the mathematical model of
blood flow coupled with the generalized plasma skimming model can be used as a
predictive tool that can estimate the delivering efficiency of drug carriers.
For the prediction, the shape parameter, $\alpha$, can be the indicator for
predicting whether the drug carriers are fast accumulated in the capillary
beds. For example, as plotted in Fig.~\ref{fig8}, when $\alpha$ is below 0.3,
the concentration of drug carriers was lower than the concentration in the
macrocirculation. Therefore, it is possible to give a guideline to enhance the
accumulation in the capillary beds by applying $\alpha$ of drug carriers. In
the current analysis, $\alpha$ = 0.3 was considered.

It must be noted that the plasma skimming model developed by Gould et al. is
capable of addressing multifurcations. The proposed method in this study is a
simple extension of the drift flux model. Hence, whilst this study used a
binary tree-like structure for studying drug carrier redistribution, this new
model is also perfectly applicable to multifurcating microvasculature.
Furthermore, by no means this study aims to provide the drug carrier specific
value of shape parameter $\alpha$ and its physical interpretation. Its
interpretation described in Section~\ref{gps} is only an example of how the
drug carriers are likely to be redistributed. Such interpretation is based on
the idea that if drug carriers have good margination, they are likely to move
with plasma after bifurcation, not RBCs in the core, and vice-versa. This shape
parameter $\alpha$ makes the proposed model to be very generalized. By
adjusting only a single parameter, the mathematical model can be easily
tailored to a specific drug carrier redistribution problem.

For the correlation between $\alpha$ and targeting efficiency, various
parametric studies \textit{in silico}, \textit{in vitro} and \textit{in vivo}
will be required. However, this study was aimed at developing a mathematical
model. To correlate the current model to a real system, the parameters for the
generalized plasma skimming model must be obtained from \textit{in vitro} and
\textit{in vivo} experiments. Also, the correlation between $M$ and $\alpha$
needs to be further investigated by conducting detailed numerical simulations with
individual RBCs, drug carriers, and blood plasma. By overcoming all these
limitations, it is hoped to utilize this new model for precisely targeting
diseased areas in microvascular networks.